# Spectroscopic signatures of many-body correlations in magic angle twisted bilayer graphene

Yonglong Xie[1], Biao Lian[2], Berthold Jäck[1], Xiaomeng Liu[1], Cheng-Li Chiu[1], Kenji Watanabe[3], Takashi Taniguchi[3], B. Andrei Bernevig[1], Ali Yazdani[1‡]

[1]*Joseph Henry Laboratories & Department of Physics, Princeton University, Princeton, NJ 08544, USA*
[2]*Princeton Center for Theoretical Science, Princeton University, Princeton, New Jersey 08544, USA*
[3]*National Institute for Material Science, 1-1 Namiki, Tsukuba 305-0044, Japan*

‡ email: yazdani@princeton.edu

**The discovery of superconducting and insulating states in magic angle twisted bilayer graphene (MATBG) [1,2] has ignited considerable interest in understanding the nature of electronic interactions in this chemically pristine material system. The phenomenological similarity of the MATBG transport properties as a function of doping with those of the high-$T_c$ cuprates and other unconventional superconductors[1,2,3] suggests the possibility that MATBG may be a highly interacting system.  However, there have not been any direct experimental evidence for strong many-body correlations in MATBG. Here we provide such evidence from using high-resolution spectroscopic measurements, as a function of carrier density, with a scanning tunneling microscopy (STM). We find MATBG to display unusual spectroscopic characteristics that can be attributed to electron-electron interactions over a wide range of doping, including when superconductivity emerges in this system. We show that our measurements cannot be explained with a mean-field approach for modeling electron-electron interaction in MATBG. The breakdown of a mean-field approach for understanding the properties of other correlated superconductors, such as cuprates, has long inspired the study of highly correlated Hubbard model[3]. We**

**show that a phenomenological extended Hubbard model cluster calculation, motivated by the nearly localized nature of the relevant electronic states of MATBG produces spectroscopic features similar to those we observe experimentally. Our findings demonstrate the critical role of many-body correlations in understanding the properties of MATBG.**

Stacking two graphene layers on top of each other results in a moiré superlattice with a periodicity depending on the angle between the layers. Near the magic angle of ~ 1º, a non-interacting continuum model of the band structure of this system predicts nearly flat low-energy valence and conduction bands[4,5]. Previously, STM studies of MATBG have visualized the moiré superlattice, identified different regions of sublattice stacking (AA and AB/BA) of graphene sheets and have resolved two peaks in the tunneling spectrum (dI/dV) associated with the large density of states (DOS) of its flat bands[9-13]. Transport studies show that the partial occupation of the two flat bands of MATBG results in a cascade of insulating phases and superconducting domes as a function of electron filling[1,2,14,15]. Considering the four-fold degeneracy (spin and valley) of the two flat bands, novel insulating phases occur at partial moiré band fillings of $\nu=n/n_0$=0, ±1/4, ±1/2, ±3/4, which suggests the predominant role of electron-electron interactions in the formation of the insulating phases. Interactions also result in the development of magnetism in MATBG[15,16], although the broken crystalline symmetry by the underlying hexagonal boron nitride (hBN) substrate may be a requirement for such phases. The observation of linear-T resistivity at high temperatures[17] may also be an indication of interactions; however, the origin of this behavior is still debated[18]. Beyond these observations and despite many different theoretical studies (for example see refs 19-23), there are many unresolved questions on the role of interactions in this system. Among these is the question of whether interactions in MATBG are not only strong when the system is insulating but also when superconductivity emerges at other doping levels[1,2,14,15]. To construct the correct model of superconducting pairing, an accurate picture of how interactions influence its low-energy excitations is required. Here we use high-resolution spectroscopy of MATBG with the STM to address these questions. Our results show that when the nearly flat valence and conduction bands are either filled or unoccupied, a non-interacting model, which includes the influence of strain and relaxation, captures the spectroscopic properties of MATBG. However, at partial band fillings, we demonstrate the strong modification of the quasi-particle spectrum from that of a non-interacting model over wide energies far exceeding that of the flat bands' bandwidth or their separation to remote bands. Without any theoretical modeling, these experimental observations illustrate that MATBG is a highly interacting problem, the physics of which cannot be captured with weak coupling theoretical approaches. The strong correlations uncovered by our experiments are key to the properties of MABTG when superconductivity emerges in this system.

We examine the properties of exposed MATBG as a function of electron density in back-gated devices (Fig. 1 a, b, see Methods for details of fabrication[24,25]) using a home-built ultra-high vacuum STM operating at 1.4K. Consistent with previous studies[9-13], STM topographies of our devices image the moiré superlattice in which the bright (dark) regions correspond to the AA (AB/BA) stacking region, where high (low) LDOS is expected (Fig. 1c). From the observed moiré lattice periodicity, we confirm the twist angle in this region of the sample to be close to the magic angle value ~ 1°. A more detailed examination of topographies also reveals the presence of strain and lattice relaxation, the information of which can be extracted from STM topographies (see Methods) and can be used to theoretically model the spectroscopic properties of MATBG when electron-electron interactions are not significant. Fig. 1d shows the dI/dV measured at the AA regions shown in Fig. 1c at a gate voltage of $V_g$= -4V, which features two sharp peaks below the Fermi level and two weaker step-like features at other energies (arrows in Fig. 1d). As expected from the continuum model and consistent with previous measurements[9-13] the two sharp peaks are associated with the van Hove singularities (vHs) of the occupied nearly flat conduction and valence bands of MATBG. However, the original non-interacting continuum model[4,5] would predict these bands to have far shaper peaks than those observed in our experiments. Including the influence of strain[8] and relaxation[6] in the continuum model results in additional dispersion of the valence and conduction bands (Fig. 1e, f), which not only better captures the width and separation of the double peaks in the tunneling spectra but also the presence of the step-like features at higher and lower energies (arrows in Fig. 1d, f). In our calculations, these step-like features are associated with the vHs of the bands remote to the flat bands. We have repeated similar local modeling of the dI/dV spectra measured at AA sites using information extracted from topographies at other locations of our devices (see Extended Data Fig. 1) and found a satisfactory description of the local spectra when disorder is weak, and when the double peaks associated with the flat bands are either below or above the chemical potential.

The breakdown of this single-particle description of the spectroscopic properties of MATBG when interactions are important becomes evident when we study the evolution of the quasi-particle spectra in our device as a function of electron density controlled by $V_g$. Fig. 2 shows dI/dV measurements on the AA region as a function of $V_g$, which spans three different regions of occupation for the two flat bands; when the flat bands are both occupied ($V_g$>-5.5 V), when they are being depleted (-53.5 V<$V_g$<-5.5 V), and after they have been depleted ($V_g$<-53.5 V). The rate of the shift of the flat bands' peaks with $V_g$ reflects the DOS at the Fermi level. Therefore, a change of slope of the lines in Fig. 2a signals a transition in band filling. The nearly vertical features signal the slow change of occupation of the flat bands with large DOS. In the range -58V<$V_g$<-53.5V, the slope change might be related to the presence of an energy gap between the flat bands and the remote bands (estimated around 15meV, roughly consistent with the calculated band structure in Fig. 1e). When the nearly flat bands are filled or fully depleted, the spectra, which are individually plotted in Fig. 2b,e, show relatively sharp double peaks at all gate voltages, the widths of which change weakly with their energy separation to the Fermi level (see Extended Data Fig. 2 and Methods). As described above, these spectra are consistent with those calculated from a non-interacting model that includes the effects of strain and relaxation. However, the most dramatic change in the quasi-particle spectra occurs when one of the flat bands begins to overlap with the Fermi level, as demonstrated by contrasting the data in Fig 2 c,d with those in Fig 2b,e. In this region, as one of the flat bands is being depleted, not only the peak associated with that band near the Fermi level develops substantial features and broadens, but surprisingly the peak associated with the other valence (conduction) band below (above) the Fermi level is also dramatically modified. Remarkably, the strong distortion of the shape of the quasi-particle spectra caused by interactions during the partial filling of the flat bands spans an energy range (30-50meV) wider than their separation to the remote bands and their apparent bandwidths, as measured when fully occupied/unoccupied. This observation demonstrates that the largest energy scale for determining the properties of MATBG at partial filling of the flat bands is set by electron-electron interactions. This signature of strong correlations occurs not just at commensurate fillings but over all the doping ranges where transport studies[1,2,14,15] have found superconductivity in this system below 1K.

To further relate our spectroscopic measurements to the transport properties of MATBG, we plot the tunneling conductance at zero energy dI/dV(0), which is a measure of the DOS at the Fermi level, as a function of $V_g$ in Fig. 3. From the changes in dI/dV(0) and the measured energies of the vHs peaks in Fig. 2, we identify the $V_g$ values corresponding to when locally we have fully depleted the conduction or valence flat bands ($n=\pm n_0$) at which point the transport measurements find evidence for a band insulator. Further assignment of occupation level within the flat bands with gate voltage is made complicated in our experiments by the presence of the STM tip (see Methods and Extended Data Fig. 3). Focusing on the gate region of $V_g$ at which we are depleting the conduction flat band, a region where the tip induced effects are minimal, we find some features in the tunneling spectra that correlate with the transport studies. Most notably, we find that at half filling of the conduction band ($n=n_0/2$), where transport measurements find the strongest insulating behavior, dI/dV(0) vanishes and a gap like feature appears in the spectrum. Recent STM studies of MATBG have reported a similar gap feature but with less than 50% of suppression of the DOS at the Fermi level[12,13]. However, we caution that this gap at half-filling is much larger (~20 times) than that observed in transport measurements and may be related to a soft gap observed at other doping levels (Fig. 2c). Interactions together with the localization of electrons either by disorder or large magnetic fields are well known to induce soft Coulomb gaps in tunneling spectroscopy[26-28]. We also identify a region close to the charge neutrality point (CNP), when the two flat band peaks are roughly symmetric about the Fermi level (green curve in Fig. 3b), where we observed a strong enhancement (~20meV) of the separation between them. This enhanced separation is another indication of the significance of interactions in MATBG[23]. Remarkably, the peaks associated with the fully occupied/unoccupied flat bands regain some of their sharpness near the CNP. The suppression of the DOS at the Fermi level near the CNP is also consistent with a recent report of insulating behavior at zero doping[15]. The apparent gap at $n=n_0/2$ and other finer features in the spectra at other rational fillings show variability between devices and may require cleaner samples to be fully established as intrinsic effects. Nevertheless, our key fundamental finding of the dramatic deviation from single-particle spectra at partial fillings is reproduceable (see Extended Data Fig. 4) and unique to magic-angle device (see Extended Data Fig. 5 for non-magic-angle device), thus provides information to discriminate between different models of interactions in MATBG.

Our first approach towards understanding the effects of electronic correlations on the spectroscopic properties of MATBG is to perform self-consistent Hartree-Fock calculations by adding the Coulomb interaction to the continuum model (See Methods). These calculations allow the possibility of interaction-induced spontaneous valley/spin polarization, without making any assumptions about the symmetry breaking in the ground state. The resulting spectra from these calculations (see Extended Data Fig. 6) show the pinning of the flat bands when being filled (vertical feature in Fig 2a) and generic broadening of the bands as well as the enhanced separation between the vHs near the CNP due to the exchange interactions. However, they fail to reproduce the abrupt distortion of the quasi-particle spectra during the partial filling of the flat bands. During the partial filling of the conduction flat band, our calculations show no discernable distortion of the valence flat band. More generally, if a mean-field order parameter is added to the bands self-consistently, there would be no reason to expect a large distortion of the un-occupied valence flat band when the occupation of the conduction flat band is slowly changed. We therefore conclude that a weak coupling mean-field picture of interactions is inadequate in producing spectra that match our findings.

Next, we consider whether the features we observe can be a consequence of the symmetry breaking in the ground state of MATBG[21]. It is instructive to contrast our findings with recent STM studies of valley polarized quantum Hall states, in which the occupation of the flat Landau level (LL) bands can be adjusted[29, 30]. These experiments show that interactions produce symmetry-broken valley polarized ground states, which induce spectral splitting at the Fermi level. However, in stark contrast to our observation, the line shape of the filled or unfilled LLs in those experiments remain unaffected even when symmetry is broken in the system. We therefore conclude that in MATBG while interactions can break the valley or spin symmetry at fractional fillings of the flat bands[15,23], such changes should only alter the states of the partially filled band, in which such symmetries are broken and would not generically alter the fully occupied or unoccupied flat bands. We attribute this behavior previously observed in quantum Hall ferromagnets to the fact that cyclotron energy is much larger than that of the interactions. By contrast, in MATBG, our incapacity to explain the features of the data within the weakly interacting model (mean-field) forces us to consider a physical picture where the effects of interactions are dominant.

The salient features of our data can be captured within a phenomenological model in which the effects of Coulomb interactions of the nearly localized states of the moiré flat bands can be studied without relying on a mean-field approximation. To motivate our phenomenological model, we note that the maximally localized Wannier orbitals corresponding to the MATBG's flat bands are shaped as three lobes symmetrically distributed around the AB/BA moiré sites, with their wavefunctions strongly localized at the three nearby AA sites[19-21]. Theoretical studies have emphasized that the shape of these Wannier functions suggests that including the nearest-neighbor Coulomb interactions is required and have constructed extended Hubbard models based on this idea. Given that the charge density is peaked on the AA sites (Fig 1c), the simplest model to consider is that of a triangular lattice with two-orbitals per site at energies $\pm\epsilon$ (no spin/valley flavor). We include a hoping between nearest sites (t) and both the on-site (U) and the nearest-neighbor ($V_0$, $V_1$) Coulomb repulsions (Fig. 4a, see Methods). To compare the results of such a model with those of our experiments, we carry out an exact diagonalization calculation for small clusters (see Methods and Extended Data Fig. 7). The resulting local spectral weight from this toy model as a function of electron filling (Fig. 4b) exhibits the same behavior as seen in the experimental results in Fig. 2b-e. When the two orbitals in the model, corresponding to the two flat bands, are fully occupied or empty, the spectra show sharp peaks (broadened by a bandwidth 6t~5meV), while at partial fillings of either orbitals (flat bands) we see substantial broadening of the two peaks. The broadening (or splitting) of the flat band peaks in our model at partial fillings is tuned by the strength of Coulomb interactions, where a value of $U/6t\sim6$ produces spectra with a broadening (~20meV) comparable to our experiments. We note that the discrete broadening (splitting) is due our use of singly degenerate orbitals to make the computation tractable, and we anticipate that restoring spin/valley flavors would further smoothen the broadened spectra with the added presence of more peaks. We also observe an interaction-enhanced energy separation between the peaks, when the chemical potential is in between the two flat bands, corresponding to the experimental results observed near the CNP. The parameters used to calculate the spectra in Fig. 4b, the onsite U (~30meV) and the nearest-neighbor V's (~5meV), have values that are consistent with recent estimates for the Coulomb interactions in magic angle moiré superlattice (with hBN underneath)[19]. Performing similar calculations on a honeycomb lattice (see Extended Data. Fig. 8), we find that our conclusions remain the same, independent of our choice of lattice structure.

Our toy model succeeds in capturing the broadening of MATBG quasi-particle spectra because it incorporates various correlated charge configurations at partial fillings, which significantly affect the electron addition/removal energies of the localized orbitals. The mean-field approach replaces these configurations with a simple average and ignores the fluctuations among them, thus results in far less complex spectral features, in contrast to those seen experimentally. More realistic extended Hubbard models with larger cluster calculations[31] can be used to more precisely compute the electron addition/removal spectra for MATBG. Most broadly, our observation that U is the largest energy scale in the problem indicates the inaccuracy of models that project the physics of electronic interactions into the two lowest flat bands. Our results indicate that MATBG is a truly many-body problem—possibly not confined to the lowest two bands. This poses a formidable theoretical challenge for finding the true ground-state for realistic models as the size of the Hilbert space, which includes valley and spin flavors, is far beyond the limit of computation with a classical computer. Regardless, our experiments and toy model calculations demonstrate that many-body correlations, at least as complex as those present in a Hubbard model are required to describe the low-energy properties of this system. Therefore, our finding establishes a more concrete connection between MATBG and high-Tc cuprates beyond the phenomenological resemblance of their transport phase diagram.

**Acknowledgements**

We acknowledge fruitful discussions with N.P. Ong, S. Wu, D. Wong, P. Sambo, L. Glazman, P. Phillips, and A. MacDonald. We also gratefully acknowledge collaborations with E. Tutuc, K. Kim, Y. Wang at the initial stages of this project. This work has been primarily supported by Gordon and Betty Moore Foundation as part of EPiQS initiative (GBMF4530) and DOE-BES grant DE-FG02-07ER46419. Other support for experimental effort has come from NSF-MRSEC programs through the Princeton Center for Complex Materials DMR-142054, NSF-DMR-1608848, ExxonMobil through Andlinger Center for Energy & the Environment at Princeton, and Princeton Catalysis Initiative. B.J. acknowledges funding of the Alexander-von-Humboldt foundation through a Feodor-Lynen postdoctoral fellowship. K.W. and T.T. acknowledge support from the Elemental Strategy Initiative conducted by the MEXT, Japan, A3 Foresight by JSPS and the CREST (JPMJCR15F3), JST. B.L. acknowledges the support by Princeton Center for Theoretical Science at Princeton University. B.A.B. acknowledges support from the Department of Energy DE-SC0016239, Simons Investigator Award, the Packard

Foundation, the Schmidt Fund for Innovative Research, NSF EAGER grant DMR-1643312, and NSF-MRSEC DMR-1420541.

**Author Contributions**

Y.X., B.J., C.L.C., X.L. and A.Y. designed and conducted the STM measurements and their analysis. X.L., Y.X. and B.J. fabricated the sample. X.L. carried out the finite element electrostatic simulation. B.L. and B.A.B. performed the theoretical calculations. K.W. and T.T. provided hBN crystals. All authors contributed to the writing of the manuscript.

**Author Information**

Reprints and permissions information is available at http://www.nature.com/ reprints. The authors declare no competing financial interests. Correspondence and requests for materials should be addressed to A.Y. (yazdani@princeton.edu).

**Figure captions:**

**Figure 1 | Non-interacting spectroscopic properties of MATBG. a,** Schematic of the STM measurement setup on MATBG devices. **b,** Optical image of the device. **c,** STM topography showing the moiré superlattice with $\theta=1.01°$. **d,** STM spectrum measured on an AA site for slight electron doping ($V_g$= -4V, $V_{set}$=200mV, $I_{set}$=120pA, $V_{mod}$=1mV). The blue and green arrows mark the step-like features. **e,** Band structure calculated using the continuum model including the effects of strain and relaxation. $\Gamma_M \Lambda_m$ is a non-high symmetry direction along which the Dirac points (locally protected by $C_{2z}T$ symmetry) are located. The black dotted line indicates the Fermi level. The blue (green) dashed line corresponds to the vHs of the first conduction (valence) remote band. **f,**

Corresponding sqrt(LDOS) (offset by -17meV) calculated on an AA site. The blue and green arrows mark the vHs of the first conduction and valence remote bands.

**Figure 2 | Breakdown of non-interacting description. a,** dI/dV measured on an AA site as a function of sample bias and gate voltage ($V_{set}$=200mV, $I_{set}$=120pA, $V_{mod}$=1mV). **b-e,** normalized dI/dV spectra at different gate voltages extracted from (**a**). The curves are shifted vertically by 0.5 each for clarity. (**c**) and (**d**) demonstrate the breakdown of the non-interacting description. The black dotted line indicates the Fermi level.

**Figure 3 | Evidence for correlated insulating states at integer fillings. a,** Conductance at the Fermi level as a function of gate voltage. Grey areas correspond to fully (un-)occupied flat bands with sharp QPs. The blue rectangles define the full filling of the two flat bands. The CNP is located inside the green region. The light yellow red rectangles is the midpoint between the edges of CNP and full fillings. The black dotted line marks the zero conductance. The asymmetry in gate responses between the conduction and valence flat bands is due to the tip band bending (see Methods). **b,** Individual spectra at different gate voltages from different shaded areas in (**a**). They are vertically shifted by 1.8 nS each for clarity. The black dotted line indicates the Fermi level.

**Figure 4 | Extended Hubbard model. a,** Schematics of a 7-site cluster two-orbital Hubbard model with on-site energy $\pm\epsilon$, hopping t, on-site Coulomb interaction U, near-neighbor interactions $V_0$ (same orbital) and $V_1$ (different orbitals). **b,** Local spectral weight computed from the exact diagonalization of a 7-site cluster Hubbard model for different filling factors with $\epsilon = 9$, t=0.75, U=30, $V_0$=5, $V_1$=3.8 (see Methods). The curves beyond $\pm n_0$ are obtained by assuming a constant DOS at the Fermi level from the remote bands. The curves are shifted vertically by 0.25 each for clarity. The double-arrow indicates the band broadening (splitting). **c,** Band broadening as a function of the ratio of the on-site Coulomb interaction U to the non-interacting band width 6t, while keeping $V_0$=U/6, $V_1$=U/7.9.

**Methods**

**Sample preparation.** The device was made using a modified 'tear and stack' technique[24,25]. We used a polypropylene carbonate (PPC) film and polydimethylsiloxane (PDMS) stack on a glass slide to first pick up a 30-40nm thick hexagonal boron nitride(hBN) flake. Then we used the van der Waals force between hBN and monolayer graphene to tear and pick up half of the graphene flake. The remaining graphene flake on the Si substrate is rotated by 1.3° and picked up. In order to expose the graphene surface, the resulting stack with PPC is transferred onto a 2nd PDMS stamp. After dissolving the PPC film in acetone, the inverted stack is placed in between pre-patterned Au contacts deposited on a $SiO_2$ wafer with a Si back gate (Fig. 1a, b). Before inserting into STM, the sample was annealed for 10hrs in ultra-high vacuum at 250 ℃. In these devices, we adjust the electron density in situ in the STM by adjusting the back-gate voltage. The 1.72° device (Extended Data Fig. 5) was fabricated by first picking up a graphite flake followed by the preparation of hBN/twisted bilayer graphene stack. The resulting stack with the PPC film was dropped off directly on pre-patterned $SiO_2$ wafer. The PPC film is removed by radiation heating in a high vacuum chamber. Finally, a 2nd graphite flake was placed to ensure the connectivity between the twisted bilayer graphene and the pre-patterned Au contact.

**STM measurements.** All measurements were performed on a homebuilt UHV STM operating at T=1.4 K, which can store two samples simultaneously. The dI/dV spectra are acquired using standard lock-in technique with a modulation voltage $V_{mod}$ (rms, f= 4 kHz) and a time constant of 5ms while keeping the feedback opened using an appropriate bias $V_{set}$ and current $I_{set}$ to stabilize the tip-sample distance.

**Estimating the effect of strain using a heterostrain model.** Due to the presence of strain, STM topographies show different moiré periodicities along different superlattice directions. We estimate the effect of strain and a more precise value of the local twist angle using a uniaxial strain model as described in ref. 12. The free parameters are the twist angle $\theta$, the strength of strain $\varepsilon$ and the direction of strain $\theta_s$. We numerically find the parameters set (θ, ε, θ_s) that best reproduces the three Moiré wavelengths from STM topography. For the first sample with $L_1$=14.04 nm, $L_2$=14.98 nm, $L_3$=12.84 nm, we find (θ=1.01°, ε=0.3%, θ_s=27°). For the second sample (see Extended Data Fig.1a), the Moiré wavelengths are $L_1$=13 nm, $L_2$=12.2 nm, $L_3$=11.8 nm and we find (θ=1.14°, ε=0.2%, θ_s=35°).

**Extended continuum model.** We modified the continuum model as described in ref. 8 to include the effect of strain. The influence of relaxation is incorporated by estimating the area ratio of AA and AB regions, which translates to the ratio of $u_{AA}/u_{AB}$ as described in ref. 6. Here we set $u_{AA}/u_{AB}$=0.83 from Figure 1c and 0.80 from Extended Data Figure 1a. The amplitude and the angle of the strain, the twist angle, the ratio of interlayer tunneling amplitude extracted from STM topography analysis are used to calculate the band structure and LDOS. The superlattice Hamiltonian contains 8 shells (146 bands per spin per valley) in our calculation. We note that the presence of uniaxial strain seen in STM topography implies the $C_{3z}$ symmetry is broken. The absence of degeneracy at $K_M$ (see Fig. 1e) is consistent with this symmetry breaking. We note that the Dirac points are still protected by $C_{2z}T$, but there is no more constraint on their locations.

**Lifetime broadening.** We extract the width of the vHs of the flat bands when they are both occupied ($V_g$>-5.5 V) and emptied ($V_g$<-53.5 V) using Gaussian fitting. The peak width as a function of the vHs energy for both conduction and valence flat bands is plotted in Extended Data Fig. 2. For fully occupied bands (Extended Data Fig. 2a), we find that the peak width follows an $E^2$ broadening with a proportional factor $\lambda$ that depends on the details of available scattering channels near the Fermi surface, as expected from the Fermi liquid theory, until the vHs energy is ~20meV below the Fermi level. For vHs energy lower than -20meV, the peak width appears to reach a plateau, 10meV for conduction flat band and 13.5meV for valence flat band. We speculate that the deviation from Fermi liquid theory possibly comes from the further distance to the Fermi level and the detailed Fermi surface when the Fermi level is inside the first conduction remote band(s). For fully emptied bands (Extended Data Fig. 2b), the peak width first decreases until the vHs energy reaches 17meV. We interpret this behavior as the suppression of electron-electron scattering channels due to the presence of an energy gap separating the valence flat band and the first valence remote band(s). The vHs energy from which the peak width starts to increase reads approximately 17meV, consistent with the gap value found in Fig. 2a. For vHs energy higher that 17meV, the peak width can be fit with an $E^2$ function with almost 30 times smaller $\lambda$, possibly due to a different Fermi surface when the Fermi level is crossing the first valence remote bands(s).

**Gate voltage asymmetry between the two flat bands.** In Fig. 3a, we find that the scale of gate voltage for the positive side of the charge neutrality (depletion of conduction flat band) is consistent with that estimated from capacitive coupling to a silicon back gate for a twist angle of 1.01°. The scale of gate voltage for the depletion of the valence flat band, i.e. negative side of the CNP, is however much smaller. This asymmetry is due to the tip-induced band bending caused by the difference in the work function between the STM tip and the sample. This band bending changes the effective doping of the region underneath the tip relative to regions further away. In this situation, the region where measurements of the DOS are performed (under the tip) is doped not only by the gate but also by the regions surrounding it—both of which respond to varying gate voltage. As a result, the density under the tip is not simply a linear function of Vg as it would be in a transport experiment. To understand this effect of tip gating, we have performed self-consistent finite element electrostatic simulation using the band structure shown in Fig.1e without considering interactions. The detailed geometry of the simulation is shown in Extended Data Fig. 3a. The results of the calculations in Extended Data Fig. 3b confirm that the density in the region under the STM tip (blue curve) is not simply a linear function of Vg. This density changes more quickly when the valence flat band overlaps with the Fermi level because of the influence of regions in the sample away from the tip. As a result, the gate voltages required to deplete the conduction and valence flat bands are different, as measured by the gate dependence of dI/dV(0) in both the model (Extended Data Fig. 3c) and measurements in Fig. 3a. The simulation in Extended Data Fig. 3c shows a 2 to 1 ratio for the difference in depletion of conduction to the valence band. This ratio in the data in Fig. 3a is 3.8 to 1; however, the exact results in the model depends on microscopic details such as the exact tip shape, work function difference, tip-sample distance. Extended Data Fig. 3d,e show the measurement on the same AA site with the same microscopic STM tip (as Fig. 3a) but with different set point conditions, which adjust the tip height. We find that the gate range for the valence flat band is sensitive to the tip height consistent with our model and the ratio of conduction to valence flat bands is reduced from that of Fig. 3a. The sensitivity of the dI(0)/dV to fractional filling of valence and conduction bands (i.e. gaps) , and the gate voltages at which they occur, also change with the height of the tip. Extended Data Fig. 3d,e also show that the regions of zero conductance near CNP depends on the tip height. Using different tip conditioning, we also find example in which the gate voltage for depleting the lowest conduction and valence bands is nearly symmetric (Extended Data Fig. 5). As noted in the manuscript, further experiments are required to associate features in dI/dV(0) with intrinsic behavior of MATBG and to correlate them with transport studies precisely. Last, in the same finite elements electrostatic simulation, we find that the tip band bending effect has negligible influence on the line shape of the flat bands.

**Self-consistent Hartree-Fock calculations.** We apply the self-consistent Hartree-Fock (HF) method to the extended continuum model in the presence of Coulomb interaction. The Coulomb interaction takes the form

$$H_I = \frac{1}{2} \sum_{\alpha,\beta,q,k,k'} \frac{V(q)}{A} \psi^\dagger_{\alpha,k+q} \psi_{\alpha,k} \psi^\dagger_{\beta,k'-q} \psi_{\beta,k'},$$

where $V(q) = 2\pi e^2/\epsilon q$ (for $q \neq 0$; for $q = 0$ one has $V(0) = 0$) is the Fourier transform of Coulomb interaction $e^2/\epsilon r$, $A$ is the area of the sample. $\psi_{\alpha,k}$ is the annihilation operator of the Dirac electron of graphene at momentum $k$, and $\alpha$ denotes the sublattice, layer, spin and valley indices. The mean field Hartree term and Fock term are given by

$$\Sigma_H = \sum_{\alpha,\beta,q,k,k'} \frac{V(q)}{A} \langle \psi^\dagger_{\alpha,k+q} \psi_{\alpha,k} \rangle \psi^\dagger_{\beta,k'-q} \psi_{\beta,k'},$$

$$\Sigma_F = -\left[ \sum_{\alpha,\beta,q,k,k'} \frac{V(q)}{A} \langle \psi^\dagger_{\alpha,k+q} \psi_{\beta,k'} \rangle \psi^\dagger_{\beta,k'-q} \psi_{\alpha,k} + h.c. \right].$$

In the extended continuum model, the momenta $k$ and $k + m_1 g_1 + m_2 g_2$ of the same spin and same valley are coupled via (multiple) momentum space hoppings, where $g_1$ and $g_2$ are the reciprocal vectors of the moiré superlattice. We assume the translation symmetry is unbroken, so $\langle \psi^\dagger_{\alpha,k} \psi_{\beta,k'} \rangle$ is nonzero only if $k - k' = m_1 g_1 + m_2 g_2$ ($m_1, m_2 \in \mathbb{Z}$) and $\alpha, \beta$ belong to the same spin and valley. The mean field HF Hamiltonian is then given by

$$H = H_0 + \Sigma_H + \Sigma_F,$$

where $H_0$ is the free Hamiltonian of the continuum model. For each fixed total filling density, we diagonalize $H$ and calculate $\Sigma_H$ and $\Sigma_F$ iteratively to find the self-consistent solution. In the calculations, we allow the 4 different spin/valley flavors to have different fillings, so that the system could spontaneously develop a flavor polarization. We keep 6 momentum shells in the continuum model (74 bands per spin per valley) for our self-consistent HF calculations. Our calculations (Extended Data Fig. 5b) show generic broadening of the flat bands independent of the fillings in contrast to the abrupt broadening seen in the experiments and possibly a gap near half filling of the conduction flat band. In addition, we find a small spontaneous spin/valley polarization (when the Fermi level is in the flat bands) with maximum amplitude near half filling of the flat bands.

**Exact diagonalization of an extended Hubbard model.** To capture the flat band broadening, we use exact diagonalization (ED) to study two single-flavor toy models of two nearly flat bands with on-site and neighboring-site interactions. The first model is defined on a triangular lattice (representing the lattice of AA stacking positions), with two orbitals per site corresponding the two flat bands, respectively. The Hamiltonian is given by

$$H = \sum_i [\epsilon(n_{1,i} - n_{2,i}) + U_1 n_{1,i} n_{2,i}] + \sum_{\langle ij \rangle} t(c^\dagger_{1,i} c_{1,j} + c^\dagger_{2,i} c_{2,j} + h.c.)$$
$$+ \sum_{\langle ij \rangle} V_0 (n_{1,i} n_{1,j} + n_{2,i} n_{2,j}) + V_1 (n_{1,i} n_{2,j} + n_{2,i} n_{1,j}),$$

where $\langle ij \rangle$ represents neighboring sites in a triangular lattice. We perform our ED for both 6-site cluster and 7-site cluster and extract out the spectral weight at various fillings. The chemical potential $\mu = dE_g(N)/dN$ is calculated from the ground state energy $E_g(N)$ at various total electron number $N$. For the 6-site ED, we assume a $2 \times 3$ triangular lattice (see Extended Data Fig. 6a) with periodic boundary conditions in both directions. For the 7-site ED (see Fig. 4a), periodic boundary condition is impossible; instead, we assume $\langle ij \rangle$ runs over all pairs of sites, so that each site still has 6 neighbors (as is true for a triangular lattice), and all the sites are equivalent. Owing to the delocalized nature of the remote bands, the interaction between the remote band and the flat bands is insignificant. When both flat bands are empty or fully occupied, and the Fermi level is in the remote band, the spectral weight only shifts in energy with respect to the electron filling. The slope is determined by the density of states of the remote band. The second model is defined on a honeycomb lattice (representing the lattice of AB and BA stacking positions, see Extended Data. Fig. 7a), where each site has one orbital, so that there are two bands in total. The Hamiltonian is given by

$$H = \sum_i \epsilon \eta_i n_i + \sum_{\langle ij \rangle} \left[ t(c_i^\dagger c_j + h.c.) + V_0 n_i n_j \right] + \sum_{\langle\langle ij \rangle\rangle} V_1 n_i n_j,$$

where $i$ runs over all the honeycomb lattice sites. $\eta_i = \pm 1$ for sublattice AB and BA, respectively, and $\epsilon$ is a staggered potential between the two sublattices. $\langle ij \rangle$ and $\langle\langle ij \rangle\rangle$ denote the nearest and next nearest neighbours, respectively. $t$ is the nearest neighbor hopping, while $V_0$ and $V_1$ are the interactions between nearest neighbor and next nearest neighbor, respectively. Since each site has only one orbital, there is no on-site interaction. Our ED calculation for this model is done for 6 unit cells, which forms a $2 \times 3$ honeycomb lattice with periodic boundary conditions. Independent of the choice of the lattice, our ED calculations show a generic broadening (splitting) of the band broadening stemming from the interplay between strong correlations and flat bands, as seen in the experiment. Adding spin/valley degeneracy would create more possible charge configurations for a given fillings, thus would further smoothen the broadened spectra with the added presence of more peaks, as can be already seen in the honeycomb calculations (see Extended Data Fig.7b).

## Data availability

The data that support the findings of this study are available from the corresponding author on reasonable request.

## Extended Data Figure Captions

**Extended Data Figure 1 | Non-interacting spectroscopic properties on another region. a,** STM topography showing the Moiré superlattice with θ=1.14°. **b,** STM spectrum measured on an AA site for slight electron doping (Vg=-10V, $V_{set}$=-200mV, $I_{set}$=500pA, $V_{mod}$=0.5mV). Blue and green arrows mark the step-like features. **c,** Band structure calculated using the continuum model including the effects of strain and relaxation. $M_M \Lambda_m$ is a non-high symmetry direction along which the Dirac points (protected by $C_{2z}T$ symmetry) are located. The black dotted line indicates the Fermi level. The blue (green) dashed line corresponds to the vHs of the first conduction (valence) remote band. **d,** Corresponding sqrt(LDOS) (offset by -28meV) on an AA site. The blue and green arrows mark the vHs of the first conduction and valence remote bands.

**Extended Data Figure 2 | Lifetime broadening. a, b,** Peak width of the vHs as a function of the vHs energy when both flat bands are filled (a) or emptied (b). The green (a) and yellow (b) curves are the fit using $p(E) = p_0 + \lambda E^2$ with $p_0 = 5.25\ meV, \lambda = 0.021\ meV^{-2}$ for (a) and with $p_0 = 5.1\ meV, \lambda = 0.00072\ meV^{-2}$. The black dotted line marks E=17 meV.

**Extended Data Figure 3 | Tip band bending effect a,** Schematics (not to scale) of the electrostatic simulation for the device geometry in our experiment. **b,** Density under and away the tip as a function of gate voltage, calculated using the geometry in (a) with z=4Å, r=0.6nm, a work function difference of 0.25V and the band structure from Fig. 1e. **c,** Density and DOS at the Fermi level as a function of gate voltage under the tip with the same parameters as in (b). **d,e,** Zero bias conductance as a function of gate

voltage with different set point conditions ($V_{set}$=+200mV, $I_{set}$=120pA for (d) and $V_{set}$=-200mV, $I_{set}$=500pA for (e)) showing different apparent gate efficiencies.

**Extended Data Figure 4 | Additional example of the breakdown of non-interacting description. a,** dI/dV spectra measured on an AA site as a function of sample bias (energy) and gate voltage on the region with θ=1.14° ($V_{set}$=-200mV, $I_{set}$=500pA, $V_{mod}$=0.5mV). **b,** normalized dI/dV spectra at different gate voltages extracted from (**a**). The curves are shifted vertically by 0.5 each for clarity. The black dotted line indicates the Fermi level.

**Extended Data Figure 5 | Density dependent spectroscopy on a non-magic angle device. a,** dI/dV measured on an AA site as a function of sample bias and gate voltage on the region with θ=1.72° ($V_{set}$=-100mV, $I_{set}$=180pA, $V_{mod}$=2mV). **b-e,** normalized dI/dV spectra at different gate voltages extracted from (**a**). The curves are shifted vertically by 0.4 each for clarity. The black dotted line indicates the Fermi level. We note that the electron and hole dopings are symmetric, which can be achieved by different tip conditioning (see Methods for discussion).

**Extended Data Figure 6 | Mean-field calculations. a, b,** normalized dI/dV spectra obtained from a non-interacting model (a) and a Hartree-Fock calculation (b) for different filling factors. The curves are each shifted vertically by 0.3 each for clarity. **c,** Individual flavor filling as a function of total filling factor indicating the presence of spontaneous spin or valley polarization near half filling of the flat bands from the Hartree-Fock calculation.

**Extended Data Figure 7 | Extended Hubbard model with 6 sites. a,** Schematics of a 6-site cluster two-orbital Hubbard model with on-site energy $\pm\epsilon$, hopping t, on-site Coulomb interaction U, near-neighbor interactions $V_0$ (same orbital) and $V_1$ (different orbitals). **b,** Local spectral weight computed from the exact diagonalization of a 6-site cluster Hubbard model for different filling factors with $\epsilon = 9$, t=0.75, U=30, $V_0$=5, $V_1$=3.8 (see Methods). The curves beyond $\pm n_0$ are obtained by assuming a constant DOS at the Fermi level from the remote bands. The curves are shifted vertically by 0.25 each for clarity. **c,** Band broadening as a function of the ratio of the on-site Coulomb interaction U to the non-interacting band width 6t, while keeping $V_0$=U/6, $V_1$=U/7.9.

**Extended Data Figure 8 | Extended Hubbard model on a honeycomb lattice. a,** Schematics of a 6-unit-cell lattice with one orbital per site, hopping t, nearest site interaction $V_0$ (different sublattices) and next-nearest site interaction $V_1$ (same lattices). **b,** Local spectral weight computed from the exact diagonalization of a 6-unit-cell lattice Hubbard model for different filling factors with $\epsilon = 8.5$, t=0.75, $V_0$=18.2, $V_1$=4.4 (see Methods). The curves beyond $\pm n_0$ are obtained by assuming a constant DOS at the Fermi level from the remote bands. The curves are shifted vertically by 0.25 each for clarity. **c,** Band broadening as a function of the ratio of the nearest site interaction $V_0$ to the non-interacting band width 3t, while keeping $V_0$=4.2*$V_1$.

Figure 1

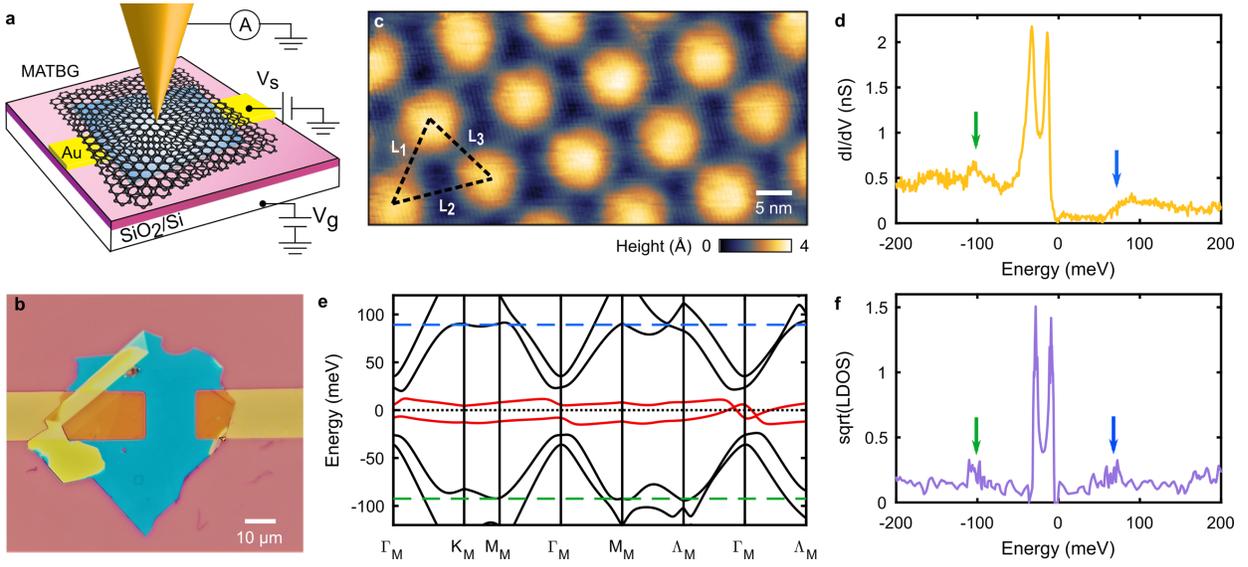

Figure 2

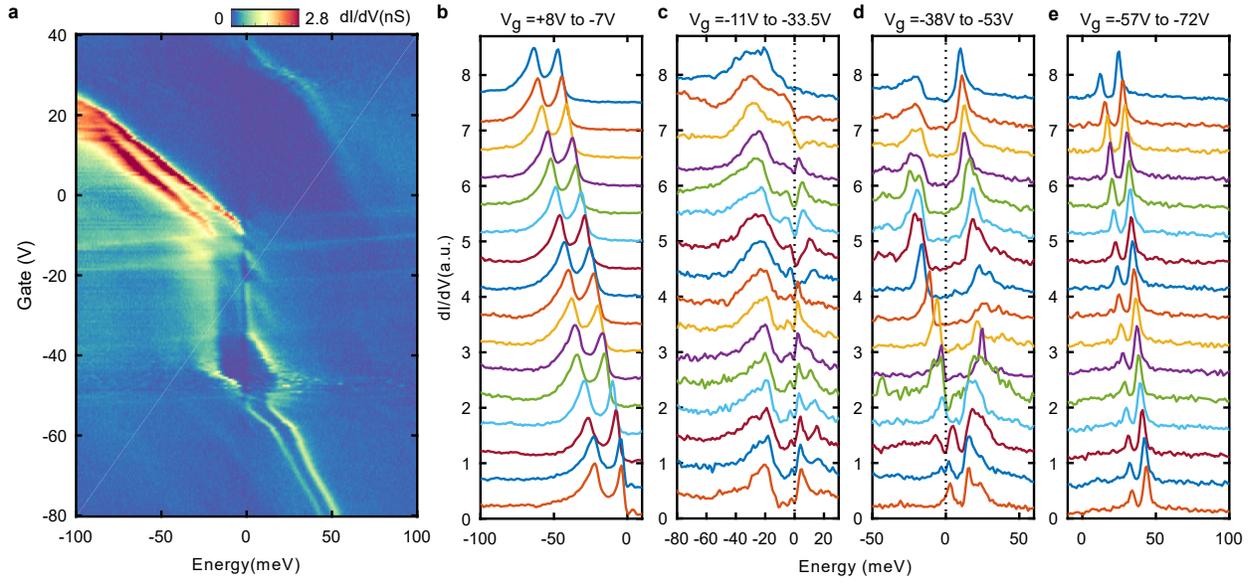

# Figure 3

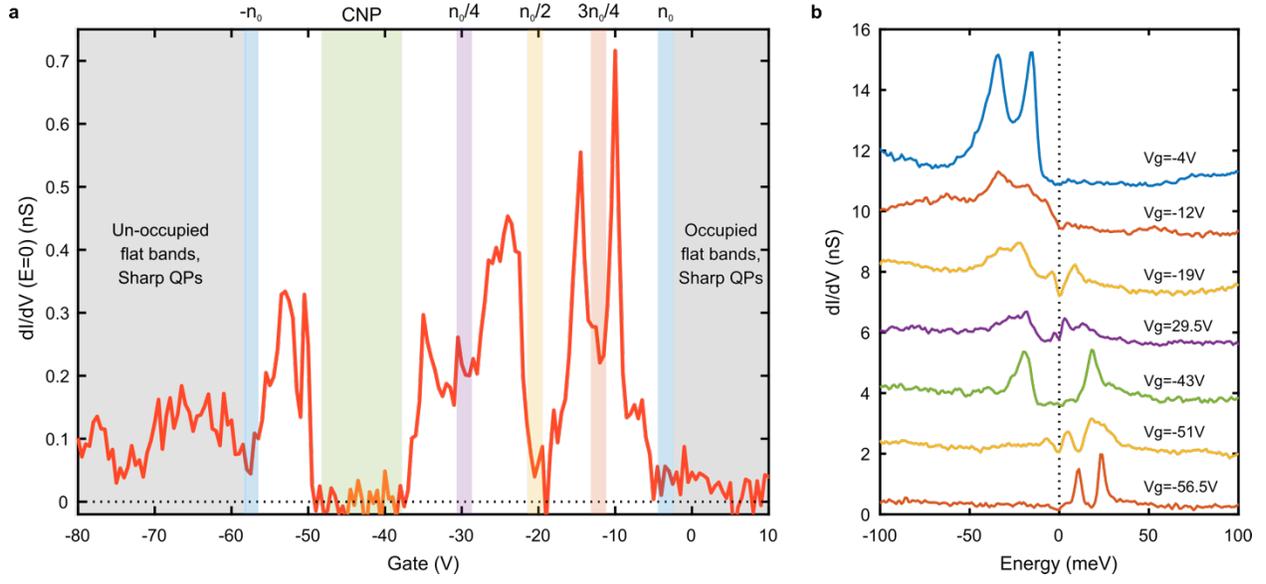

Figure 4

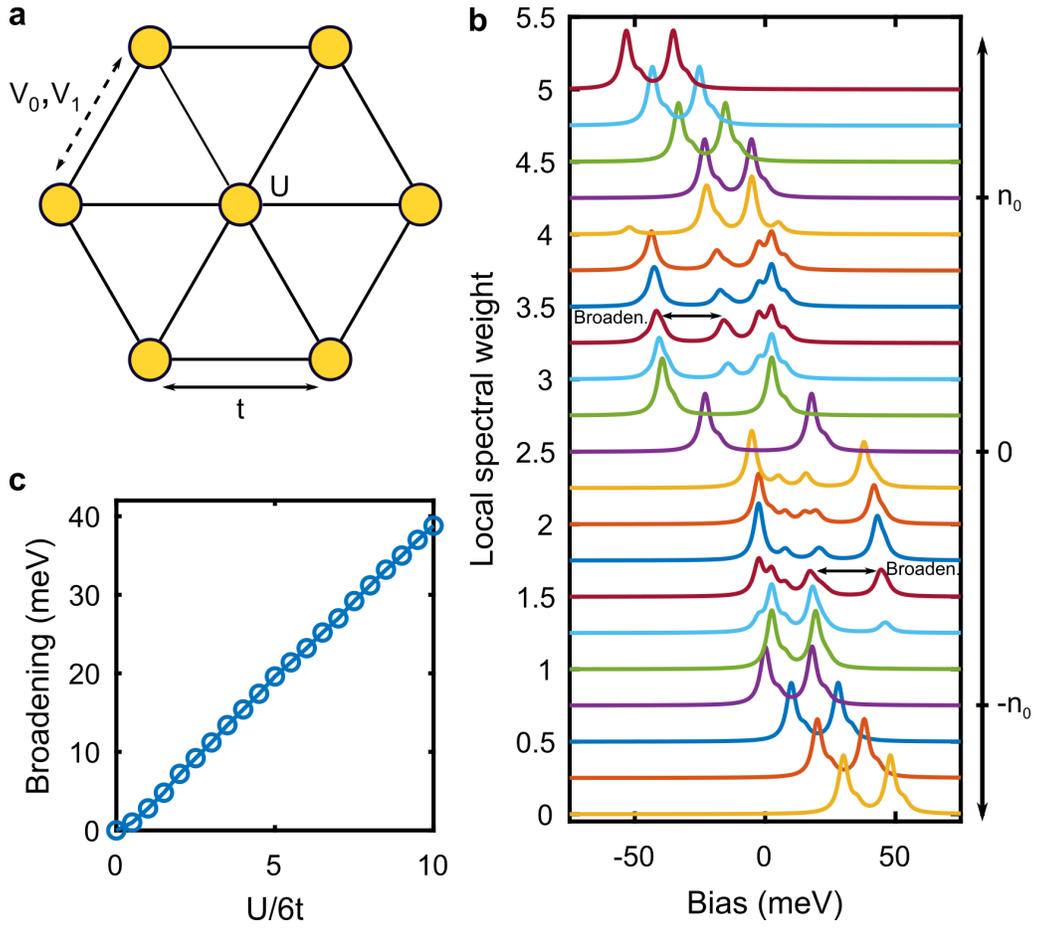

## Extended Data Figure 1

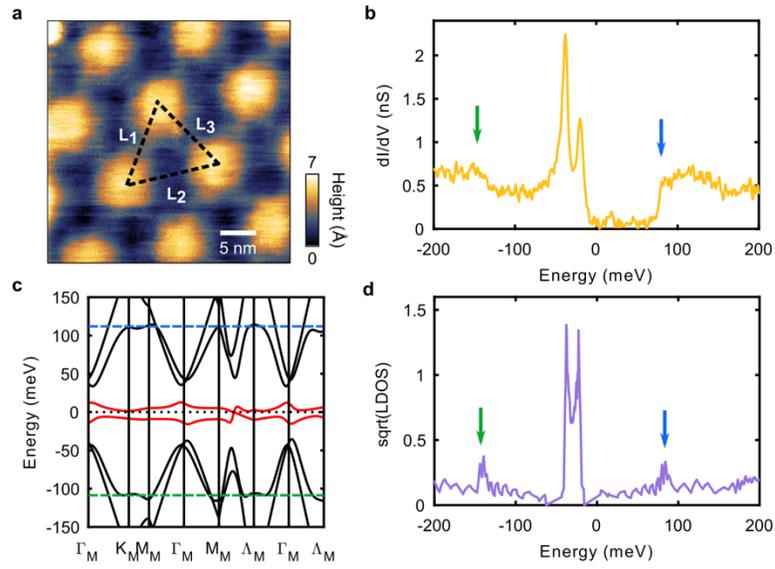

# Extended Data Figure 2

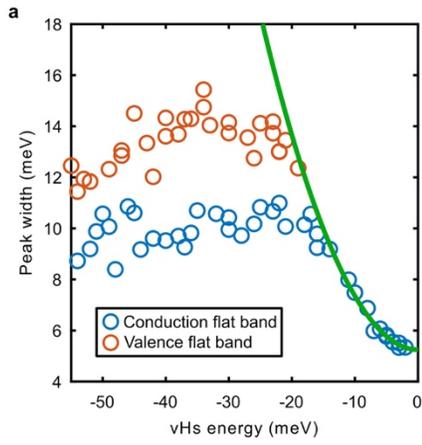 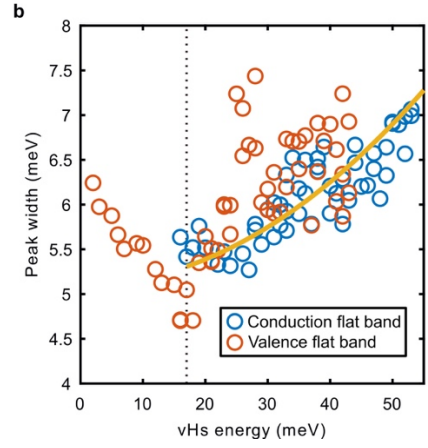

# Extended Data Figure 3

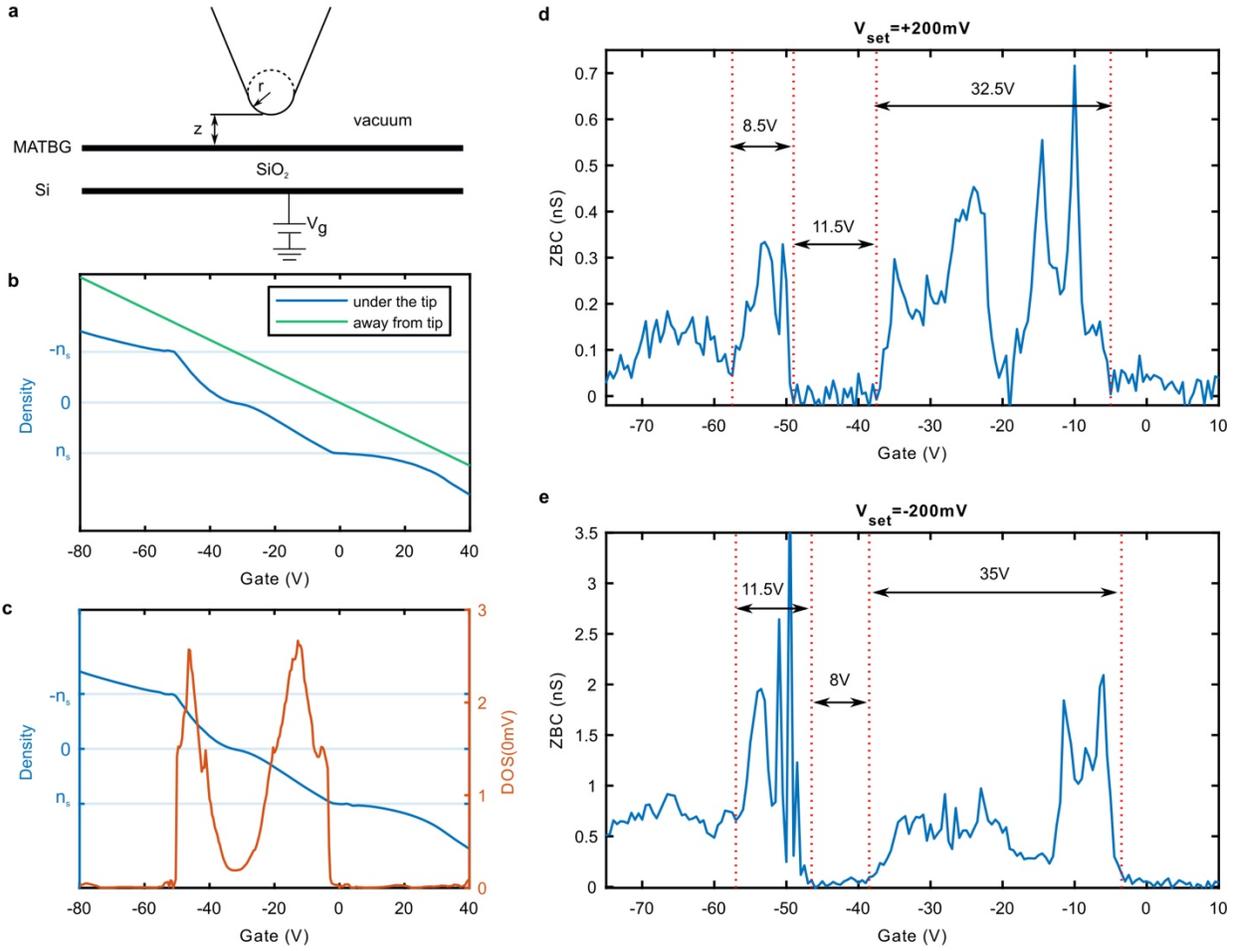

# Extended Data Figure 4

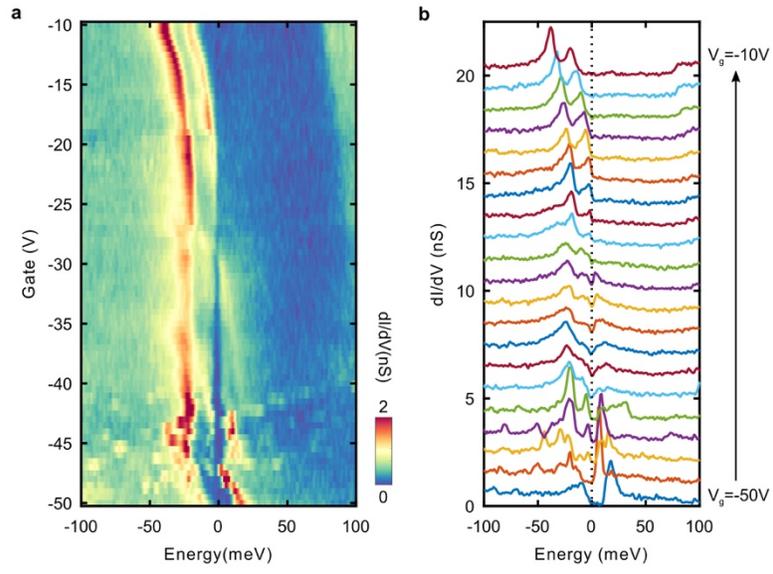

## Extended Data Figure 5

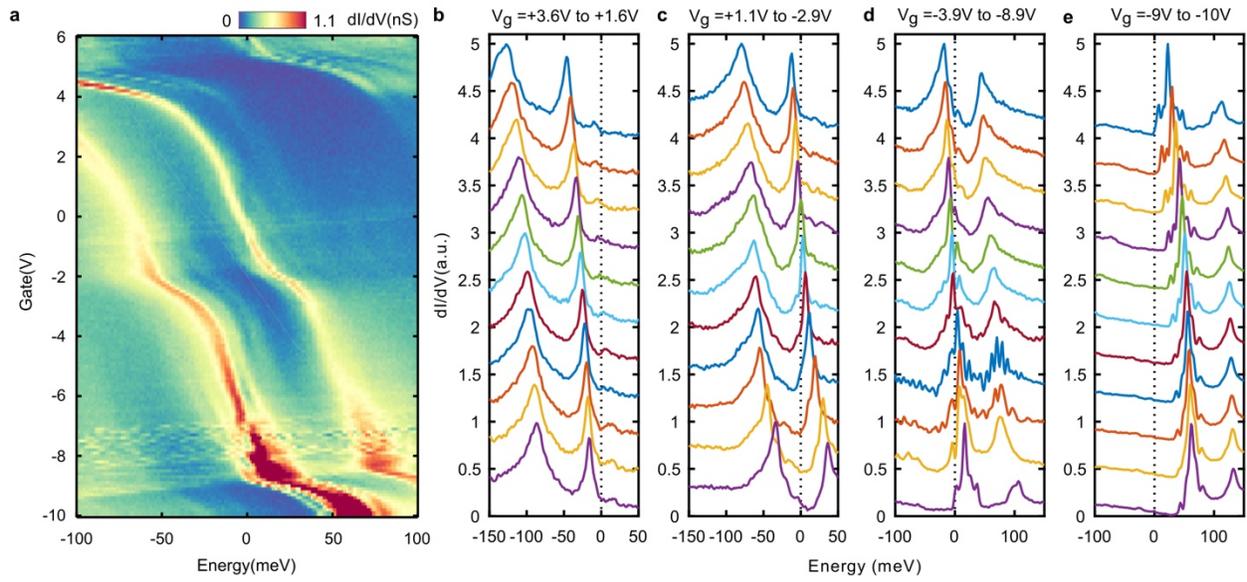

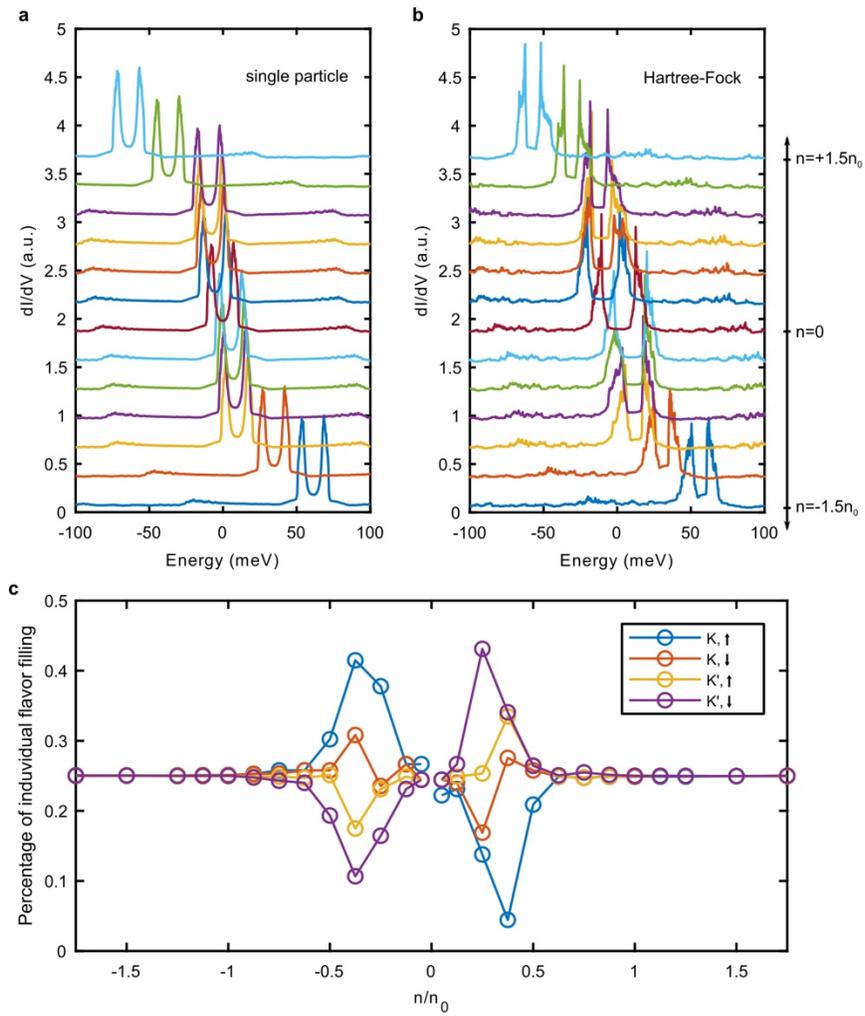

## Extended Data Figure 7

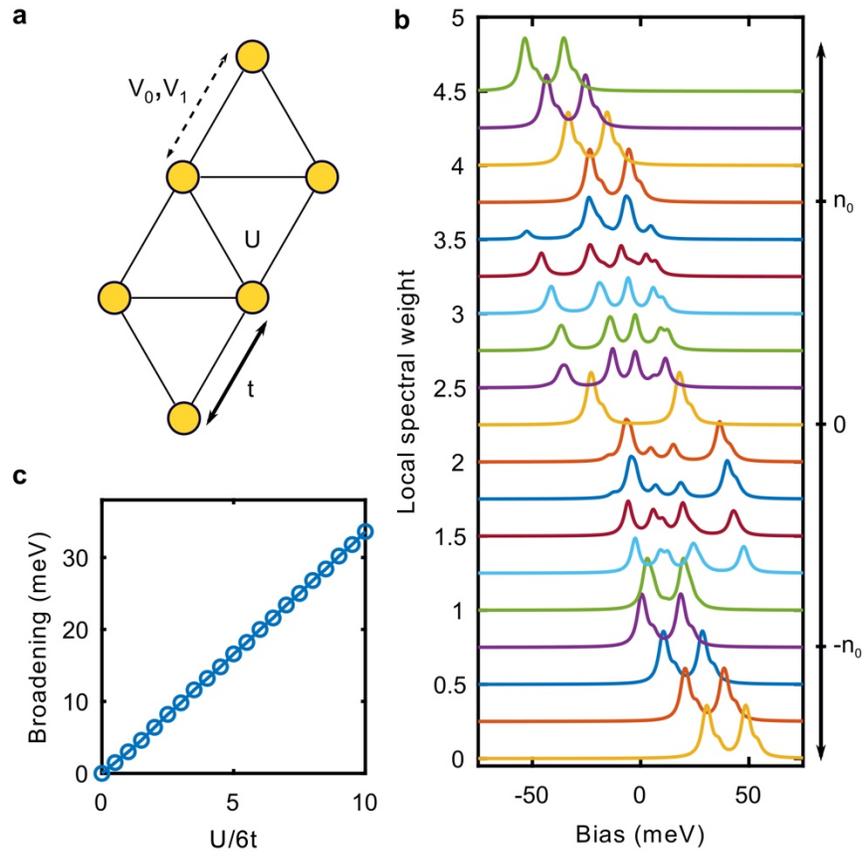

# Extended Data Figure 8

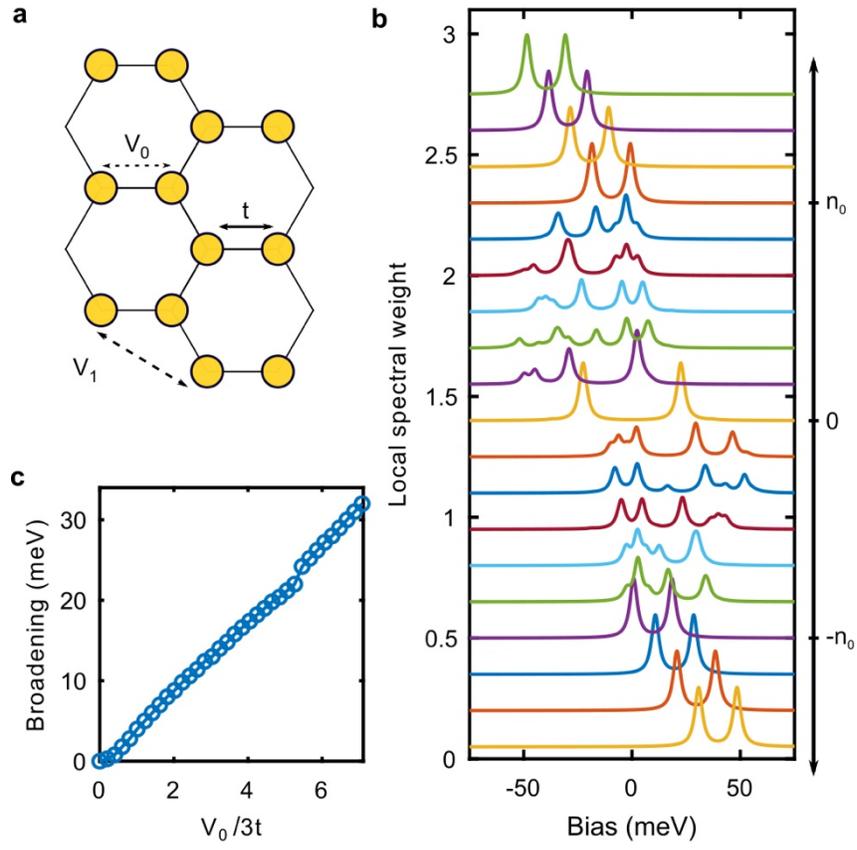